\documentclass[final,1p,times,twocolumn]{elsarticle}
\usepackage{graphicx}
\usepackage{amssymb}
\usepackage{amsthm}
\usepackage{lineno}

\begin{document}
\begin{frontmatter}
\title{Nonequilibrium Keldysh Formalism for Interacting Leads --
Application to Quantum Dot Transport Driven by Spin Bias}

\author[a,b]{Yuan Li\fnref{label2}}
\ead{liyuan@hdu.edu.cn}
\author[a,c]{M. B. A. Jalil}
\author[c]{Seng Ghee Tan}

\address[a]{Department of Electronic and Computer Engineering,
Information Storage Materials Laboratory, National University of
Singapore, 1 Engineering Drive 3, Singapore 117576}
\address[b]{Department of Physics, Hangzhou Dianzi University, Hangzhou
310018, P. R. China}
\address[c]{Data Storage Institute, DSI Building, 5

Engineering Drive 1, National University of Singapore, Singapore
117608,Singapore}
 \fntext[label2]{Corresponding author. Address:
Information Storage Materials Laboratory, Electrical and Computer
Engineering Department, National University of Singapore, 4
Engineering Drive 3, Singapore 117576, Singapore.}

\begin{abstract}
The conductance through a mesoscopic system of interacting electrons
coupled to two adjacent leads is conventionally derived via the
Keldysh nonequilibrium Green's function technique, in the limit of
noninteracting leads [see Y. Meir \emph{et al.}, Phys. Rev. Lett.
\textbf{68}, 2512 (1991)]. We extend the standard formalism to cater
for a quantum dot system with Coulombic interactions between the
quantum dot and the leads. The general current expression is
obtained by considering the equation of motion of the time-ordered
Green's function of the system. The nonequilibrium effects of the
interacting leads are then incorporated by determining the
contour-ordered Green's function over the Keldysh loop and applying
Langreth's theorem. The dot-lead interactions significantly increase
the height of the Kondo peaks in density of states of the quantum
dot. This translates into two Kondo peaks in the spin differential
conductance when the magnitude of the spin bias equals that of the
Zeeman splitting. There also exists a plateau in the charge
differential conductance due to the combined effect of spin bias and
the Zeeman splitting. The low-bias conductance plateau with sharp
edges is also a characteristic of the Kondo effect. The conductance
plateau disappears for the case of asymmetric dot-lead interaction.
\end{abstract}

\end{frontmatter}

\section{Introduction}
With the advancement in nanotechnology, a small number of
interacting electrons can be confined in a quantum dot, to mimic the
behavior of impurity atoms in a metal, thus enabling the study of
many-body correlations between electrons. One such many-body
phenomenon is the Kondo effect, the discovery of which in quantum
dot systems \cite{Ng1,Glazman}, has generated tremendous theoretical
\cite{Meir2,Pustilnik,Sun,Martinek} and experimental
\cite{Gordon2,Sasaki} interest. Early works on the Kondo effect had
largely focused on the effect of charge bias which applied to the
two leads. Later, attentions have shifted to the Kondo effects of
the systems including ferromagnetic leads
\cite{Sergueev,Zhang,Pasupathy,Utsumi} and the leads with spin
accumulation \cite{Koop}. Spin accumulation is the nonequilibrium
split in the chemical potentials of spin-up and spin-down electrons.
Accordingly, a spin bias can be realized experimentally, for
instance, by controlling the spin accumulation at the biased
contacts between ferromagnetic and nonmagnetic leads, and can in
turn, generate a pure spin current without any accompanying charge
current \cite{Kato,Valenzuela}. Recent experimental progress has
opened new possibilities for the study of spin-bias-induced
transport in strongly correlated systems
\cite{Bao,Swirkovicz,Kobayashi}. However the effect of the spin bias
on the Kondo effect remains largely unexplored. More importantly,
the Kondo effect in the presence of Coulombic interactions between
the quantum dot and two adjacent leads has not been studied.

Such dot-lead interactions warrant an investigation owing to the
increasing influence of leads with the reduction in the lead-dot
separation. The nonequilibrium (bias-driven) transport through a
interacting mesoscopic system is usually modeled by the Keldysh
nonequilibrium Green's function (NEGF) method. The standard Keldysh
NEGF method for a central quantum dot system coupled to two biased
leads, was first developed by Meir \emph{et al.} \cite{Meir0}, and
widely applied thereafter \cite{Jauho,Nardelli,Sun2,Li,Kumar}.

In this paper, we derive the general analytical expression for the
current in a quantum dot system with interacting leads. The dot-lead
Coulombic interactions are then incorporated by determining the
contour-ordered Green's function over the Keldysh loop and applying
Langreth's theorem. Based on the current formula, and the retarded
Green's function of the quantum dot system, we study the
low-temperature spin transport and the Kondo physics induced by a
spin bias, in the presence of dot-lead Coulombic interactions. Two sharp
peaks occur in the density of states in the Kondo regime. This translates
into Kondo peaks occur in the spin differential
conductance when the magnitude of the spin bias becomes equal to that of
the Zeeman split of the quantum dot energy levels. There also exists a
conductance plateau in the charge differential conductance at low bias, due to
the combined effect of the spin bias and the Zeeman splitting. Since the Kondo effect
primarily arises from intra-dot Coulomb interactions between electrons of opposite spins,
the Kondo peaks and hence the position of the conductance peaks and
plateau are largely unaffected by the strength of the lead-dot Coulombic interactions.
The conductance plateau disappears when the symmetry between the
dot-lead interactions in the left and right junctions is broken.

The organization of the rest of the paper is as follows. In
Sec.~\ref{model}, the Hamiltonian of the system is introduced, and
the current formula and corresponding Green's functions are derived.
In Sec.~\ref{numerical}, we present the results of our numerical
calculation of density of states and the differential conductance in
the presence of the dot-lead Coulombic interactions. Finally, a
brief summary is given in Sec.~\ref{summary}.

\section{Model and formulation}\label{model}
\label{sec:model} The Hamiltonian of the quantum dot system can be
written as $H=H_c+H_{cen}+H_T+H_{ld}$. The first term is the
Hamiltonian of the contacts
\begin{eqnarray}
H_c=\sum_{k\sigma \alpha\in \mathrm{L,R}}\varepsilon_{\alpha k
\sigma}c_{\alpha k\sigma}^{\dag}c_{\alpha k\sigma},
\end{eqnarray}
while the Hamiltonian of the central region is described by:
\begin{eqnarray}
H_{cen}=\sum_{\sigma}\varepsilon_\sigma
d_{\sigma}^{\dag}d_{\sigma}+Ud_{\uparrow}^{\dag}d_{\uparrow}d_{\downarrow}^{\dag}d_{\downarrow},
\end{eqnarray}
where $c_{\alpha k\sigma}^{\dag}$ ($c_{\alpha k\sigma}$) and
$d_{\sigma}^{\dag}$ ($d_{\sigma}$) are the creation (annihilation)
operators of an electron with spin $\sigma(=\uparrow,\downarrow)$ in
the lead $\alpha(=\mathrm{L,R})$ and the quantum dot, respectively.
The second term in $H_{cen}$ is the Anderson term where $U$ refers
to the on-site Coulombic repulsion, while $\varepsilon_{\alpha k
\sigma}$ and $\varepsilon_\sigma$ are the energy levels of the two
leads and the quantum dot, respectively. For simplicity, we consider
only a single pair of levels with energies
$\varepsilon_{\sigma}=\varepsilon_0+\sigma\Delta \varepsilon/2$. The
third term represents tunneling coupling between the leads and the
central region:
\begin{eqnarray}
H_T=\sum_{k\sigma \alpha\in \mathrm{L,R}}(t_{\alpha
\sigma}c^{\dag}_{\alpha k\sigma}d_{\sigma}+\mathrm{H.C.}),
\end{eqnarray}
where $t_{\alpha \sigma}$ refers to the tunneling coupling constant.

In the conventional NEGF method, the interactions in the leads are
disregarded so that the semi-infinite leads can be described by the
simple noninteracting Green's function (see Meir \emph{et al.}
\cite{Meir0}). In this paper, however, we incorporate the Coulombic
interaction between the charges on the quantum dot and the two
leads, which may be significant given the proximity of the two
regions. This interaction term in the Hamiltonian $H_{ld}$ is given
by\cite{Sade}
\begin{eqnarray}
H_{ld}=\sum_{\alpha k \sigma \sigma'} I_{\alpha k}d_\sigma^\dag
d_\sigma c_{\alpha k\sigma'}^{\dag}c_{\alpha k\sigma'},
\end{eqnarray}
where $I_{\alpha k}$ denotes the interaction strength. We first
derive the general formula of the current by considering the
equation of motion method, and the contour-ordered Green's function
over the Keldysh loop, as will be shown explicitly below.

The current from the lead $\alpha$ through barrier to the central
region can be calculated from the time evolution of the occupation
number operator of the lead $\alpha$:
\begin{eqnarray}\label{current1}
J_{\alpha\sigma}=-e\langle\dot{N}_{\alpha\sigma}\rangle=\frac{2e}{\hbar}\mathrm{Re}\big\{\sum_k
t_{\alpha\sigma}G_{\alpha k\sigma}^<(t,t)\big\},
\end{eqnarray}
where $N_{\alpha\sigma}=\sum_k c_{\alpha k\sigma}^{\dag}c_{\alpha
k\sigma}$ and the lesser Green's function is defined as
\begin{eqnarray}
G_{\alpha k\sigma}^<(t,t')\equiv i\langle c_{\alpha
k\sigma}^{\dag}(t')d_\sigma(t)\rangle.
\end{eqnarray}
According to the equation of motion satisfied by the time-ordered
Green's function $G_{\alpha k\sigma}^t$, namely,
\begin{eqnarray}\label{motion1}
&&-i\frac{\partial}{\partial t'}G_{\alpha
k\sigma}^t(t,t')=\varepsilon_{\alpha k \sigma}G_{\alpha
k\sigma}^t(t,t')+t_{\alpha\sigma}^\ast
G_{\sigma}^t(t,t')\nonumber\\
&&\hspace{5mm}+\sum_{\sigma''}I_{\alpha k}(-i)\langle \mathrm{T}
\{d_\sigma (t)d_{\sigma''}^\dag(t') d_{\sigma''}(t')c_{\alpha
k\sigma}^{\dag}(t')\}\rangle,
\end{eqnarray}
where we define the central region time-ordered Green's function
$G_{\sigma}^t(t,t')=-i\langle \mathrm{T}\{d_{\sigma}(t)
d_{\sigma}^\dag(t')\}\rangle$ with $\mathrm{T}$ being the
time-ordering operator. Note that the Coulombic interaction between
the quantum dot and the leads results in the last term in
Eq.~(\ref{motion1}). This additional term prevents the closure of
the equation of motion satisfied by $G_{\alpha k\sigma}^t$ except by
utilizing certain approximations. We adopt the Hartree-Fock
approximation to deal with this term and obtain a closed set of
equations. Using the Hartree-Fock approximation, the last term can
be expressed as
\begin{eqnarray*}
-i\langle T \{d_\sigma (t)d_{\sigma''}^\dag(t')
d_{\sigma''}(t')c_{\alpha k\sigma}^{\dag}(t')\}\rangle \simeq
n_{\sigma''}(-i)\langle T \{d_\sigma (t)c_{\alpha
k\sigma}^{\dag}(t')\}\rangle,
\end{eqnarray*}
here $n_{\sigma''}=\langle d_{\sigma''}^\dag d_{\sigma''}\rangle$.
Thus, the equation of motion of Eq.~(\ref{motion1}) can be rewritten
as
\begin{eqnarray}
\left[-i\frac{\partial}{\partial t'}-\varepsilon_{\alpha k
\sigma}-A_{\alpha k}\right]G_{\alpha
k\sigma}^t(t,t')=t_{\alpha\sigma}^\ast G_{\sigma}^t(t,t'),
\end{eqnarray}
where $A_{\alpha k}=n I_{\alpha k }$ with
$n=n_\uparrow+n_\downarrow$. Thus, the Green's function $G_{\alpha k
\sigma}^<(t,t')$ can be derived through adopting Langreth's theorem,
namely
\begin{eqnarray}
G_{\alpha k \sigma}^<(t,t')=\int dt_1 t_{\alpha \sigma}^\ast
\big[G_\sigma^r(t,t_1)g_{\alpha k \sigma}^<(t_1,t')
+G_{\sigma}^<(t,t_1)g_{\alpha k \sigma}^a(t_1,t')\big],
\end{eqnarray}
where the time-dependent Green's function of the leads for the
uncoupled system with the modified energy level $\varepsilon_{\alpha
k \sigma}'$ are
\begin{eqnarray}
g_{\alpha k \sigma}^<(t, t')
&=&if_{\alpha\sigma}(\varepsilon_{\alpha k
\sigma}')\exp\big[-i\varepsilon_{\alpha k \sigma}'(t-t')\big],\\
g_{\alpha k \sigma}^{(r,a)}(t, t') &=&\mp i\theta(\pm t\mp
t')\exp\big[-i\varepsilon_{\alpha k \sigma}'(t-t')\big],
\end{eqnarray}
where $f_{\alpha\sigma}(\varepsilon_{\alpha k
\sigma}')=\big\{\exp\big[(\varepsilon_{\alpha k \sigma}'-\mu_{\alpha
\sigma})/k_BT\big]+1\big\}^{-1}$. Substituting the Green's function
$G_{\alpha k \sigma}^<(t,t')$ into Eq.~(\ref{current1}), we can
obtain the formula of the current
\begin{eqnarray}
J_{\alpha \sigma}&=&\frac{ie}{\hbar}\int
\frac{d\epsilon}{2\pi}\Gamma_{\alpha\sigma}(\epsilon)\big\{G_\sigma^<(\epsilon+A_\alpha(\epsilon))\nonumber\\
&&+f_{\alpha\sigma}(\epsilon+A_\alpha(\epsilon))\big[G^r_\sigma(\epsilon+A_\alpha(\epsilon))
-G^a_\sigma(\epsilon+A_\alpha(\epsilon))\big]\big\},
\end{eqnarray}
where $\Gamma_{\alpha\sigma}=2\pi\sum_k|
t_{\alpha\sigma}|^2\delta(\epsilon-\varepsilon_{\alpha k \sigma})$
is the linewidth function and $A_{\alpha}(\epsilon)=n I_\alpha
(\epsilon)$. The Green's function $G^{r,<}_\sigma(\epsilon)$ are the
Fourier transformations of $G^{r,<}_\sigma(t)$ with
$G^{r}_{\sigma}(t)
=-i\theta(t)\langle\{d_\sigma(t),d_\sigma^\dag(0)\}\rangle\equiv
\langle\langle d_\sigma(t)|d_\sigma^\dag(0)\rangle\rangle^r$,
$G^{<}_\sigma(t) =i\langle d^\dag_\sigma(0) d_\sigma(t)\rangle\equiv
\langle\langle d_\sigma(t)|d_\sigma^\dag(0)\rangle\rangle^<$. For
the case of proportionate coupling to the leads, i.e.,
$\Gamma_{L\sigma}=\lambda\Gamma_{R\sigma}$, the formula of the
current can be simplified. Defining the current $J=xJ_{L}-(1-x)J_R$
with $J_\alpha=J_{\alpha\uparrow}+J_{\alpha\downarrow}$, we can
obtain the simplified formula of the charge current, namely
\begin{eqnarray}\label{current2}
J&=&\frac{e}{\hbar}\sum_\sigma\int
d\epsilon\left[f_{L\sigma}(\epsilon+A_\alpha(\epsilon))-f_{R\sigma}(\epsilon+A_\alpha(\epsilon))\right]\nonumber\\
&&\hspace{10mm}\times\frac{\Gamma_{L\sigma}(\epsilon)\Gamma_{R\sigma}(\epsilon)}{\Gamma_{L\sigma}(\epsilon)+\Gamma_{R\sigma}(\epsilon)}
\left[-\frac{1}{\pi}\mathrm{Im}G^r_\sigma(\epsilon+A_\alpha(\epsilon))\right].
\end{eqnarray}
Thus, finally, the current can be expressed solely in terms of the
retarded Green's function.

We now proceed to solve the retarded Green's function
$G^{r}_{\sigma}(\epsilon)$. The standard equation of motion
technique yields the following general relation
\begin{eqnarray*}
\epsilon \langle\langle \hat{F}_1|\hat{F}_2\rangle\rangle^r=\langle
\{\hat{F}_1,\hat{F}_2\}\rangle+\langle\langle \left[\hat{F}_1,
H\right]|\hat{F}_2\rangle\rangle^r,
\end{eqnarray*}
where $\hat{F}_1$ and $\hat{F}_2$ are arbitrary operators. Note that
the energy $\epsilon$ includes a infinitesimal imaginary part
$i0^+$. By considering the model Hamiltonian, the equation of motion
of the Green's function $G_\sigma^r(t)$ can thus be written as
\begin{eqnarray}\label{green1}
(\epsilon-\varepsilon_\sigma)\langle\langle
d_\sigma|d_\sigma^\dag\rangle\rangle^r&=&1+\sum_{\alpha k
\sigma'}I_{\alpha k}\langle\langle d_\sigma c_{\alpha k
\sigma'}^\dag c_{\alpha k
\sigma'}|d_{\sigma}^\dag\rangle\rangle^r\nonumber\\
&&+\sum_{\alpha k}t_{\alpha\sigma}^\ast\langle\langle c_{\alpha k
\sigma}|d_\sigma^\dag\rangle\rangle^r+U\langle\langle d_\sigma
d_{\bar{\sigma}}^\dag
d_{\bar{\sigma}}|d_{\sigma}^\dag\rangle\rangle^r.
\end{eqnarray}
The above equation of motion generates three new Green's functions,
i.e., $\langle\langle d_\sigma c_{\alpha k \sigma'}^\dag c_{\alpha k
\sigma'}|d_{\sigma}^\dag\rangle\rangle^r$, $\langle\langle d_\sigma
d_{\bar{\sigma}}^\dag
d_{\bar{\sigma}}|d_{\sigma}^\dag\rangle\rangle^r$ and
$\langle\langle c_{\alpha k \sigma}|d_\sigma^\dag\rangle\rangle^r$.
We then consider the respective equations of motions of these
Green's functions:
\begin{eqnarray}\label{green2}
(\epsilon-\varepsilon_{\alpha k \sigma})\langle\langle c_{\alpha k
\sigma}|d_\sigma^\dag\rangle\rangle^r
=t_{\alpha\sigma}\langle\langle
d_\sigma|d_\sigma^\dag\rangle\rangle^r +\sum_{\sigma'}I_{\alpha k
}\langle\langle d_{\sigma'}^\dag d_{\sigma'}c_{\alpha k
\sigma}|d_\sigma^\dag\rangle\rangle^r.
\end{eqnarray}
\begin{eqnarray}\label{equation1}
&&(\epsilon-\varepsilon_{\alpha k \sigma})\langle\langle
d_{\sigma'}^\dag d_{\sigma'}c_{\alpha k
\sigma}|d_\sigma^\dag\rangle\rangle^r\nonumber\\
&&=\sum_{\alpha' k'}t_{\alpha' \sigma'}\langle\langle c_{\alpha' k'
\sigma'}^\dag c_{\alpha k
\sigma}d_{\sigma'}|d_\sigma^\dag\rangle\rangle^r+t_{\alpha\sigma}\langle\langle
d_{\sigma'}^\dag
d_{\sigma'}d_{\sigma}|d_\sigma^\dag\rangle\rangle^r\nonumber\\
&&\hspace{5mm}+\sum_{\alpha' k'}t_{\alpha' \sigma'}^\ast
\langle\langle d_{\sigma'}^\dag c_{\alpha' k' \sigma'}c_{\alpha
k\sigma}|d_\sigma^\dag\rangle\rangle^r+\sum_{\sigma''}I_{\alpha
k}\langle\langle d_{\sigma''}^\dag d_{\sigma''}d_{\sigma'}^\dag
d_{\sigma'}c_{\alpha k \sigma}|d_\sigma^\dag\rangle\rangle^r,
\end{eqnarray}
\begin{eqnarray}\label{equation2}
&&(\epsilon-\varepsilon_\sigma-U)\langle\langle d_\sigma
d_{\bar{\sigma}}^\dag
d_{\bar{\sigma}}|d_{\sigma}^\dag\rangle\rangle^r\nonumber\\
&&=n_{\bar{\sigma}}+\sum_{\alpha'
k'}t_{\alpha'\bar{\sigma}}\langle\langle c_{\alpha' k'
\bar{\sigma}}^\dag d_\sigma d_{\bar{\sigma}}
|d_{\sigma}^\dag\rangle\rangle^r+\sum_{\alpha'
k'}t_{\alpha'\bar{\sigma}}^\ast\langle\langle d_\sigma
d_{\bar{\sigma}}^\dag c_{\alpha' k'
\bar{\sigma}}|d_{\sigma}^\dag\rangle\rangle^r\nonumber\\
&&\hspace{5mm}+\sum_{\alpha' k'}t_{\alpha'\sigma}^\ast\langle\langle
d_{\bar{\sigma}}^\dag d_{\bar{\sigma}} c_{\alpha' k'
\sigma}|d_{\sigma}^\dag\rangle\rangle^r+\sum_{\alpha' k'
\sigma'}I_{\alpha' k'}\langle\langle d_\sigma d_{\bar{\sigma}}^\dag
d_{\bar{\sigma}} c_{\alpha' k' \sigma'}^\dag c_{\alpha' k'
\sigma'}|d_{\sigma}^\dag\rangle\rangle^r,
\end{eqnarray}
\begin{eqnarray}\label{equation3}
&&(\epsilon+\varepsilon_{\alpha'k'\bar{\sigma}}-\varepsilon_{\bar{\sigma}}-\varepsilon_{\sigma}-U)
\langle\langle c_{\alpha' k' \bar{\sigma}}^\dag d_\sigma
d_{\bar{\sigma}}
|d_{\sigma}^\dag\rangle\rangle^r\nonumber\\
&&=\sum_{\alpha'' k''}t_{\alpha''\bar{\sigma}}^\ast\langle\langle
c_{\alpha' k'\bar{\sigma}}^\dag d_\sigma c_{\alpha''
k''\bar{\sigma}}|d_{\sigma}^\dag\rangle\rangle^r-\sum_{\alpha''
k''}t_{\alpha''\sigma}^\ast\langle\langle c_{\alpha'
k'\bar{\sigma}}^\dag d_{\bar{\sigma}} c_{\alpha''
k''\sigma}|d_{\sigma}^\dag\rangle\rangle^r\nonumber\\
&&+t_{\alpha'\bar{\sigma}}^\ast\langle\langle d_\sigma
d_{\bar{\sigma}}^\dag d_{\bar{\sigma}}
|d_{\sigma}^\dag\rangle\rangle^r+\sum_{\alpha''k''\sigma''}2
I_{\alpha''k''}\langle\langle c_{\alpha'k'\bar{\sigma}}^\dag
d_\sigma d_{\bar{\sigma}} c_{\alpha''k''\sigma''}^\dag
c_{\alpha''k''\sigma''} |d_{\sigma}^\dag\rangle\rangle^r,
\end{eqnarray}
\begin{eqnarray}\label{equation4}
&&(\epsilon-\varepsilon_{\alpha'k'\bar{\sigma}}+\varepsilon_{\bar{\sigma}}-\varepsilon_{\sigma})
\langle\langle d_\sigma d_{\bar{\sigma}}^\dag c_{\alpha' k'
\bar{\sigma}}|d_{\sigma}^\dag\rangle\rangle^r\nonumber\\
&&=-\langle
c_{\alpha'k'\bar{\sigma}}d_{\bar{\sigma}}^\dag\rangle-U\langle\langle(d_{\bar{\sigma}}
d_{\bar{\sigma}}^\dag+d_{\sigma}^\dag d_{\sigma})d_\sigma
d_{\bar{\sigma}}^\dag c_{\alpha' k' \bar{\sigma}}
|d_{\sigma}^\dag\rangle\rangle^r\nonumber\\
&&\hspace{5mm}+\sum_{\alpha''
k''}t_{\alpha''\bar{\sigma}}\langle\langle c_{\alpha'
k'\bar{\sigma}}c_{\alpha'' k''\bar{\sigma}}^\dag
d_{\sigma}|d_{\sigma}^\dag\rangle\rangle^r+\sum_{\alpha''
k''}t_{\alpha''\bar{\sigma}}^\ast\langle\langle c_{\alpha'
k'\bar{\sigma}} c_{\alpha'' k''\sigma} d_{\bar{\sigma}}^\dag
|d_{\sigma}^\dag\rangle\rangle^r\nonumber\\
&&\hspace{5mm}+t_{\alpha'\bar{\sigma}}\langle\langle
d_{\bar{\sigma}}d_{\sigma}d_{\bar{\sigma}}^\dag
|d_{\sigma}^\dag\rangle\rangle^r+\sum_{\sigma''}I_{\alpha'k'}
\langle\langle c_{\alpha'k'\bar{\sigma}} d_{\sigma''}^\dag
d_{\sigma''} d_{\sigma}d_{\bar{\sigma}}^\dag
|d_{\sigma}^\dag\rangle\rangle^r.
\end{eqnarray}
To obtain a closed set of equations from the above relations, we
adopt a decoupling approximation based on the following rules
\cite{Sun,Meir,Huag}: $\langle YX\rangle=0$ and $\langle\langle
YX_1X_2|d_\sigma^\dag\rangle\rangle^r\approx\langle
X_1X_2\rangle\langle\langle Y|d_\sigma^\dag\rangle\rangle^r$, where
$X$ and $Y$ represent the operators of leads and the quantum dot. We
shall also assume that higher-order spin-correlations terms can be
neglected. With these approximations,
Eq.~(\ref{equation1})-Eq.~(\ref{equation4}) simplify to
\begin{eqnarray}\label{equation5}
&&(\epsilon-\varepsilon_{\alpha k \sigma})\langle\langle
d_{\sigma'}^\dag d_{\sigma'}c_{\alpha k
\sigma}|d_\sigma^\dag\rangle\rangle^r\nonumber\\
&&\hspace{5mm}=t_{\alpha\sigma}\big[f_{\alpha\sigma}(\varepsilon_{\alpha
k \sigma})\langle\langle d_\sigma|d_\sigma^\dag\rangle\rangle^r
\delta_{\sigma\sigma'}+\langle\langle d_{\sigma}
d_{\bar{\sigma}}^\dag
d_{\bar{\sigma}}|d_\sigma^\dag\rangle\rangle^r\delta_{\sigma'\bar{\sigma}}\big],
\end{eqnarray}
\begin{eqnarray}\label{equation6}
&&(\epsilon+\varepsilon_{\alpha'k'\bar{\sigma}}-\varepsilon_{\bar{\sigma}}-\varepsilon_{\sigma}-U)
\langle\langle c_{\alpha' k' \bar{\sigma}}^\dag d_\sigma
d_{\bar{\sigma}}
|d_{\sigma}^\dag\rangle\rangle^r\nonumber\\
&&\hspace{5mm}=t_{\alpha'\bar{\sigma}}^\ast\big[\langle\langle
d_\sigma d_{\bar{\sigma}}^\dag
d_{\bar{\sigma}}|d_{\sigma}^\dag\rangle\rangle^r-f_{\alpha'\bar{\sigma}}(\varepsilon_{\alpha'
k'\bar{\sigma}})\langle\langle
d_\sigma|d_\sigma^\dag\rangle\rangle^r\big],
\end{eqnarray}
\begin{eqnarray}\label{equation7}
&&(\epsilon-\varepsilon_{\alpha'k'\bar{\sigma}}+\varepsilon_{\bar{\sigma}}-\varepsilon_{\sigma})
\langle\langle  d_\sigma d_{\bar{\sigma}}^\dag c_{\alpha' k'
\bar{\sigma}}
|d_{\sigma}^\dag\rangle\rangle^r\nonumber\\
&&\hspace{5mm}=t_{\alpha'\bar{\sigma}}\big[\langle\langle d_\sigma
d_{\bar{\sigma}}^\dag
d_{\bar{\sigma}}|d_{\sigma}^\dag\rangle\rangle^r-f_{\alpha'\bar{\sigma}}(\varepsilon_{\alpha'
k'\bar{\sigma}})\langle\langle
d_\sigma|d_\sigma^\dag\rangle\rangle^r\big],
\end{eqnarray}
after ignoring all of the higher-order terms. Substituting
Eq.~(\ref{equation5}), Eq.~(\ref{equation6}) and
Eq.~(\ref{equation7}) into Eq.~(\ref{equation2}), we then obtain the
following:
\begin{eqnarray}\label{equation8}
(\epsilon-\varepsilon_\sigma-U-\Sigma_{0\sigma}-\Sigma_{1\sigma})\langle\langle
d_\sigma d_{\bar{\sigma}}^\dag
d_{\bar{\sigma}}|d_{\sigma}^\dag\rangle\rangle^r=
n_{\bar{\sigma}}-\Sigma_{2\sigma}\langle\langle d_\sigma
|d_{\sigma}^\dag\rangle\rangle^r.
\end{eqnarray}
In the above, the self-energies $\Sigma_{0\sigma, 1\sigma,2\sigma}$
are defined as
\begin{eqnarray*}
&&\Sigma_{0\sigma}=\sum_{\alpha'k'}\frac{|t_{\alpha'\sigma}|^2}{\epsilon-\varepsilon_{\alpha'k'\sigma}},\\
&&\Sigma_{i\sigma}=\sum_{\alpha'k'}A_{\alpha'k'\bar{\sigma}}^{(i)}|t_{\alpha'\bar{\sigma}}|^2\left[\frac{1}
{\epsilon-\varepsilon_{\alpha'k'\bar{\sigma}}-\varepsilon_\sigma+\varepsilon_{\bar{\sigma}}}
+\frac{1}
{\epsilon+\varepsilon_{\alpha'k'\bar{\sigma}}-\varepsilon_\sigma-\varepsilon_{\bar{\sigma}}-U}\right],
\end{eqnarray*}
where $i=1,2$, $A_{\alpha'k'\bar{\sigma}}^{(1)}=1$ and
$A_{\alpha'k'\bar{\sigma}}^{(2)}=f_{\alpha'\bar{\sigma}}(\varepsilon_{\alpha'k'\bar{\sigma}})$.

Subsequently, by substituting Eq.~(\ref{green2}),
Eq.~(\ref{equation5}) and Eq.~(\ref{equation8}) into
Eq.~(\ref{green1}), we obtain a relation involving the Green's
function $G^r_\sigma(\epsilon)=\langle\langle d_\sigma|
d_\sigma^\dag\rangle\rangle^r$:
\begin{eqnarray}
(\epsilon-\varepsilon_\sigma-\Sigma_{3\sigma}-\Sigma_{4\sigma})\langle\langle
d_\sigma| d_\sigma^\dag\rangle\rangle^r
=1+(U+\Sigma_{5\sigma})\langle\langle d_\sigma d_{\bar{\sigma}}^\dag
d_{\bar{\sigma}}| d_\sigma^\dag\rangle\rangle^r,
\end{eqnarray}
where
\begin{eqnarray*}
&&\Sigma_{3\sigma}=\sum_{\alpha k \sigma'}I_{\alpha k
}f_{\alpha\sigma'}(\varepsilon_{\alpha k\sigma'}),
~~~\Sigma_{4\sigma}=\sum_{\alpha
k}\frac{|t_{\alpha\sigma}|^2}{\epsilon-\varepsilon_{\alpha k
\sigma}}\left[1+\frac{I_{\alpha
k}f_{\alpha\sigma}(\varepsilon_{\alpha k
\sigma})}{\epsilon-\varepsilon_{\alpha k\sigma}}\right],\\
&&\Sigma_{5\sigma}=\sum_{\alpha
k}\frac{|t_{\alpha\sigma}|^2I_{\alpha k
}}{(\epsilon-\varepsilon_{\alpha k \sigma})^2}.
\end{eqnarray*}
Finally we can obtain the analytic form of the required Green's
function, namely,
\begin{eqnarray}\label{equation9}
&&G_\sigma^r(\epsilon)\equiv\langle\langle d_\sigma|
d_\sigma^\dag\rangle\rangle^r\nonumber\\
&&\hspace{2mm}=\frac{\epsilon-\varepsilon_\sigma-U-
\Sigma_{0\sigma}-\Sigma_{1\sigma}+(U+\Sigma_{5\sigma})n_{\bar{\sigma}}}{(\epsilon-\varepsilon_\sigma-\Sigma_{3\sigma}-\Sigma_{4\sigma})
(\epsilon-\varepsilon_\sigma-U-\Sigma_{0\sigma}-\Sigma_{1\sigma})+(U+\Sigma_{5\sigma})\Sigma_{2\sigma}}.
\end{eqnarray}

In the absence of any interaction between the quantum dot and the
two leads, two of the self-energy terms reduce to zero, i.e.,
$\Sigma_{3\sigma}=\Sigma_{5\sigma}=0$, while
\begin{eqnarray*}
\Sigma_{4\sigma}=\Sigma_{0\sigma}=\sum_{\alpha
k}\frac{|t_{\alpha\sigma}|^2}{\epsilon-\varepsilon_{\alpha k
\sigma}}.
\end{eqnarray*}
Eq. (\ref{equation9}) is the Green's function of a quantum dot
system with Coulombic interactions within the dot, and between the
dot and the two leads. For simplicity, we assume the case of strong
intra-dot Coulomb interaction (i.e., the infinite-U limit), for
which the Green's function $G_\sigma^r(\epsilon)$ reduces to:
\begin{eqnarray}\label{equation10}
G_\sigma^r(\epsilon)=\langle\langle d_\sigma|
d_\sigma^\dag\rangle\rangle^r=\frac{1-n_{\bar{\sigma}}}{\epsilon-\varepsilon_\sigma-\Sigma_{2\sigma}'
-\Sigma_{3\sigma}-\Sigma_{4\sigma}},
\end{eqnarray}
where the new self energy $\Sigma_{2\sigma}'$ is defined as
\begin{eqnarray}\label{equation9a}
\Sigma_{2\sigma}'=\sum_{\alpha'k'}\frac{f_{\alpha'\bar{\sigma}}(\varepsilon_{\alpha'k'\bar{\sigma}})
|t_{\alpha'\bar{\sigma}}|^2}
{\epsilon-\varepsilon_{\alpha'k'\bar{\sigma}}-\varepsilon_\sigma+\varepsilon_{\bar{\sigma}}}.
\end{eqnarray}
Obviously, the self-energies terms $\Sigma_{3\sigma}$ and
$\Sigma_{4\sigma}$ represent the effects of the interaction between
the quantum dot and the two leads. Note that the Green's function is
dependent on the occupation number $n_\sigma$ which is given by the
formula $n_\sigma=-i\int (d\epsilon/2\pi)G_\sigma^<(\epsilon)$. The
lesser Green's function $G_\sigma^<(\epsilon)$ in the formula may be
obtained by adopting certain approximation, for instance, the
noncrossing approximation\cite{Meir2} and the ansatz
method\cite{Ng}. Alternatively, one can directly calculate the
integral $\int d\epsilon G_\sigma^<(\epsilon)$ via the following
relation:
\begin{eqnarray}\label{nsigma}
n_\sigma=-\int \frac{d\epsilon}{\pi}\mathrm{Im }G_\sigma^r
\frac{\Gamma_{L\sigma}f_{L\sigma}(\epsilon)+\Gamma_{R\sigma}f_{R\sigma}(\epsilon)}{\Gamma_{L\sigma}+\Gamma_{R\sigma}}.
\end{eqnarray}
Thus, from Eqs. (\ref{equation10}) and (\ref{nsigma}), one can
evaluate $G^r_\sigma(\epsilon)$ and $n_\sigma$ self-consistently.
Finally, the converged value of $n_\sigma$ is then used to calculate
the charge current via Eq.~(\ref{current2}).
\begin{figure}[t]
\begin{center}
\includegraphics[width=6cm]{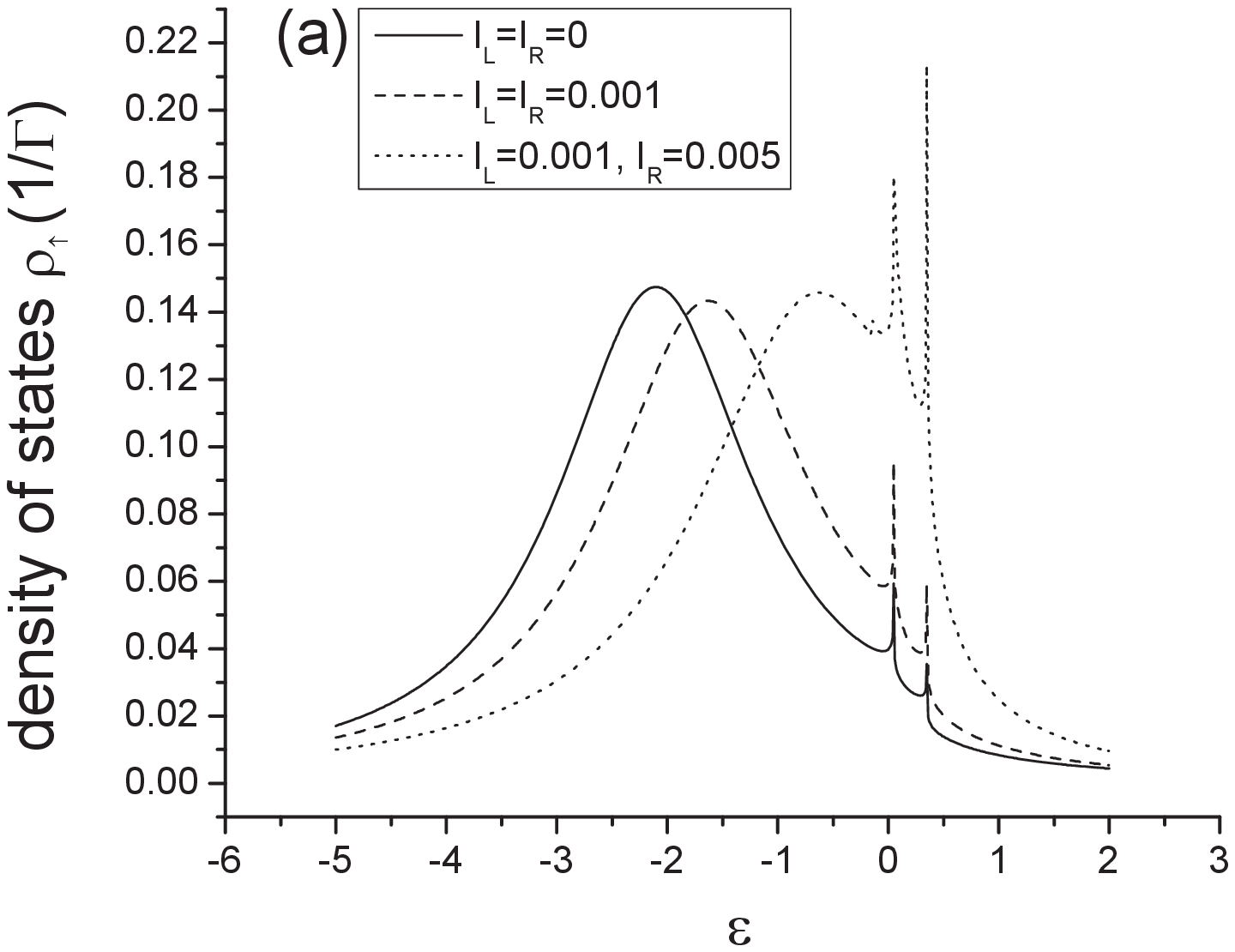}
\includegraphics[width=6cm]{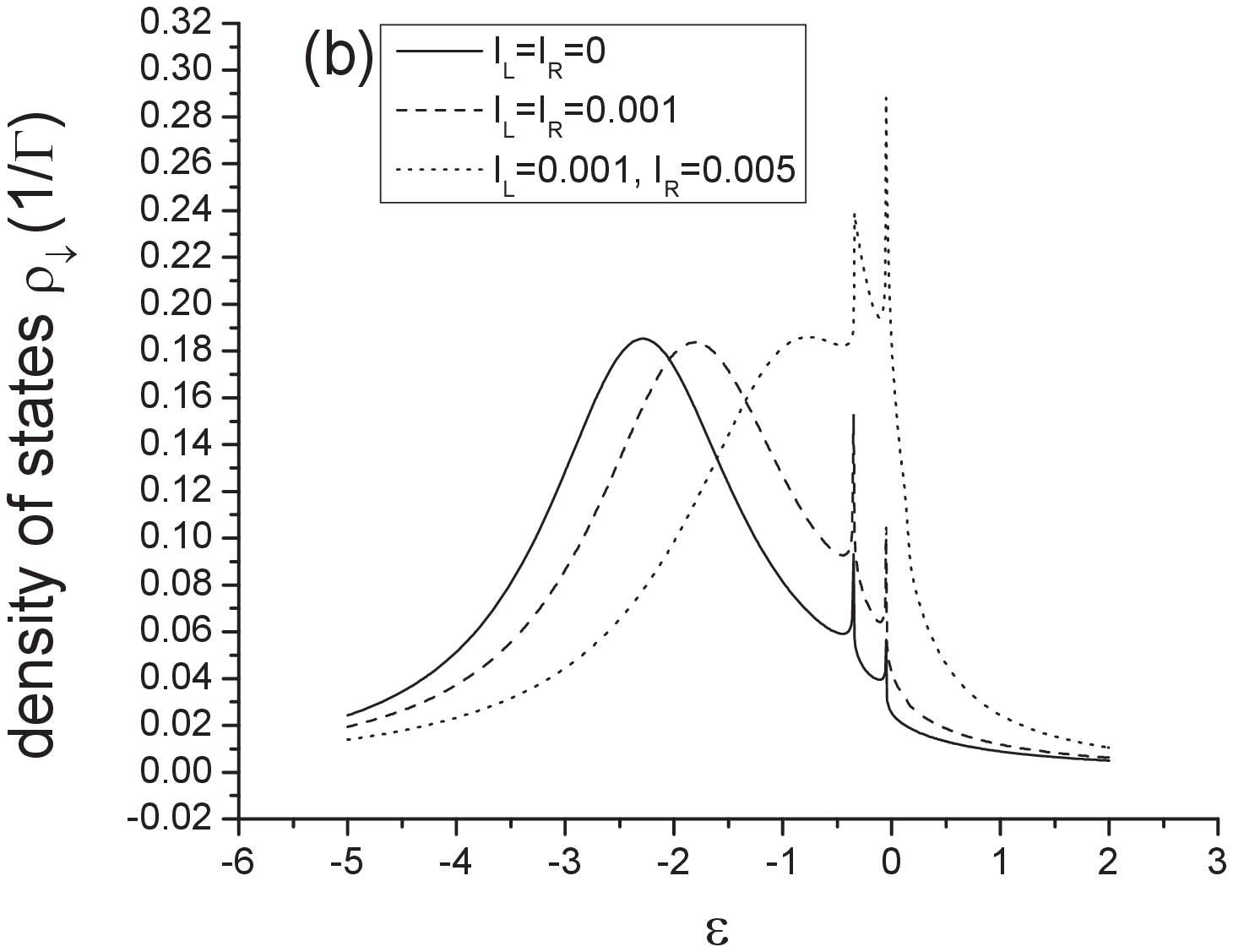}
\caption{\label{fig:midu1}(a) and (b) are schematic diagrams of the
density of states $\rho_\sigma (\epsilon)$ for a quantum dot
symmetrically coupled to two leads with Lorentzian linewidth of
$2D$. The quantum dot has two spin states with energies
$\varepsilon_\uparrow=-2.9$, $\varepsilon_\downarrow=-3.1$ and an
on-site interaction $U\rightarrow \infty$. The linewidth is chosen
to be $D=500$ and the temperature is $T=0.005$. The spin bias is
$V_s=0.3$ and the chemical potentials are
$\mu_{L\uparrow}=-\mu_{L\downarrow}=V_s/2$ and
$\mu_{R\downarrow}=-\mu_{R\uparrow}=V_s/2$. Solid, dashed and dotted
curves correspond to interaction parameters of $I_L=I_R=0$,
$I_L=I_R=0.001$ and $I_L=0.001, I_R=0.005$, respectively.}
 \end{center}
\end{figure}
\section{Numerical results}\label{numerical}
Having derived the nonequilibrium transport for the general case of
a central region coupled to interacting leads, we investigate the
spin transport through a quantum dot system with a spin bias $V_s$
applied at the two leads, namely
$\mu_{L\uparrow}=-\mu_{L\downarrow}=V_s/2$ and
$\mu_{R\downarrow}=-\mu_{R\uparrow}=V_s/2$. Our focus is to analyze
the effect of the lead-dot Coulomb interaction on the spin-transport
properties. Firstly, we start to study density of states of the
quantum dot according to the relation $\rho_\sigma
(\epsilon)=-(1/\pi)\mathrm{Im} G^r_\sigma(\epsilon)$. In the
following numerical calculations, we assume that the quantum dot
symmetrically couples to two leads with Lorentzian linewidth of
$2D$, namely
$\Gamma_{L\sigma}(\epsilon)=\Gamma_{R\sigma}(\epsilon)=\gamma_0D^2/2(\epsilon^2+D^2)$,
with $\gamma_0=1$ as the unit of energy and $D=500$. As for the
dot-lead interaction, we adopt a flat-band profile, i.e.,
$I_\alpha(\epsilon)=I_\alpha\theta(D-| \epsilon|)$.

\begin{figure}[!t]
\begin{center}
\includegraphics[width=5cm]{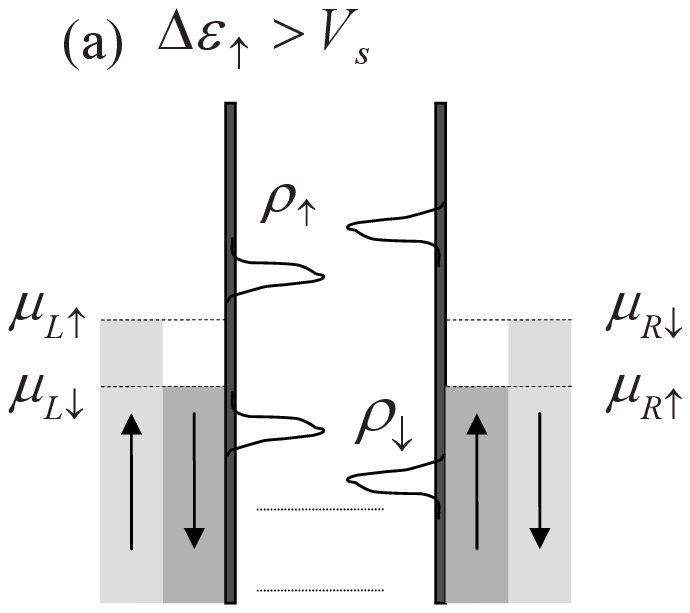}
\includegraphics[width=5cm]{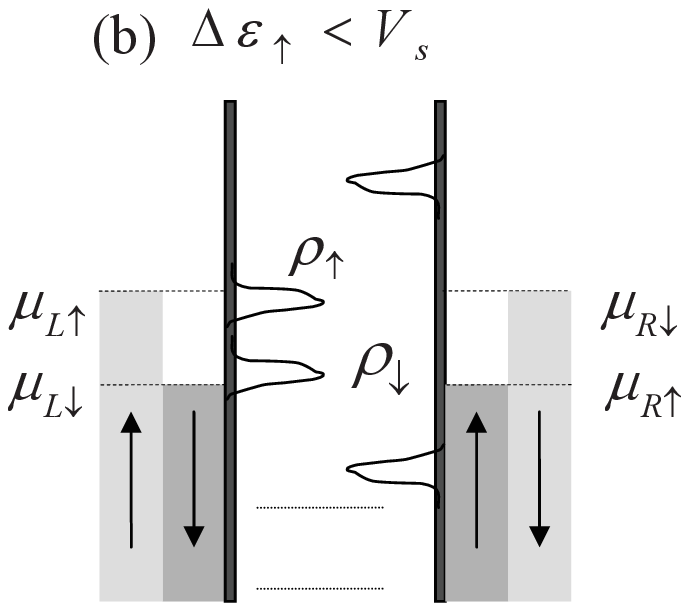}
\includegraphics[width=6cm]{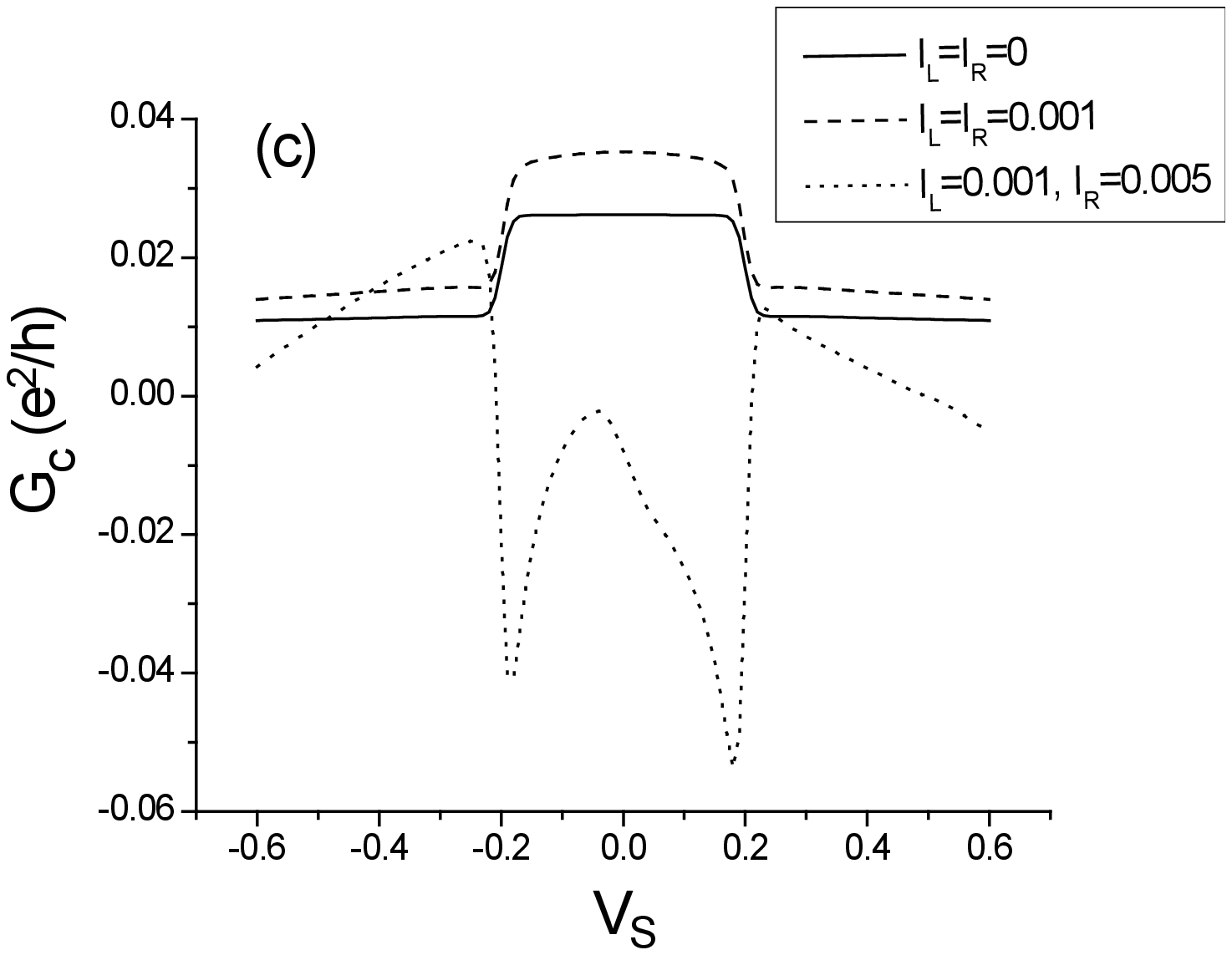}
\includegraphics[width=6cm]{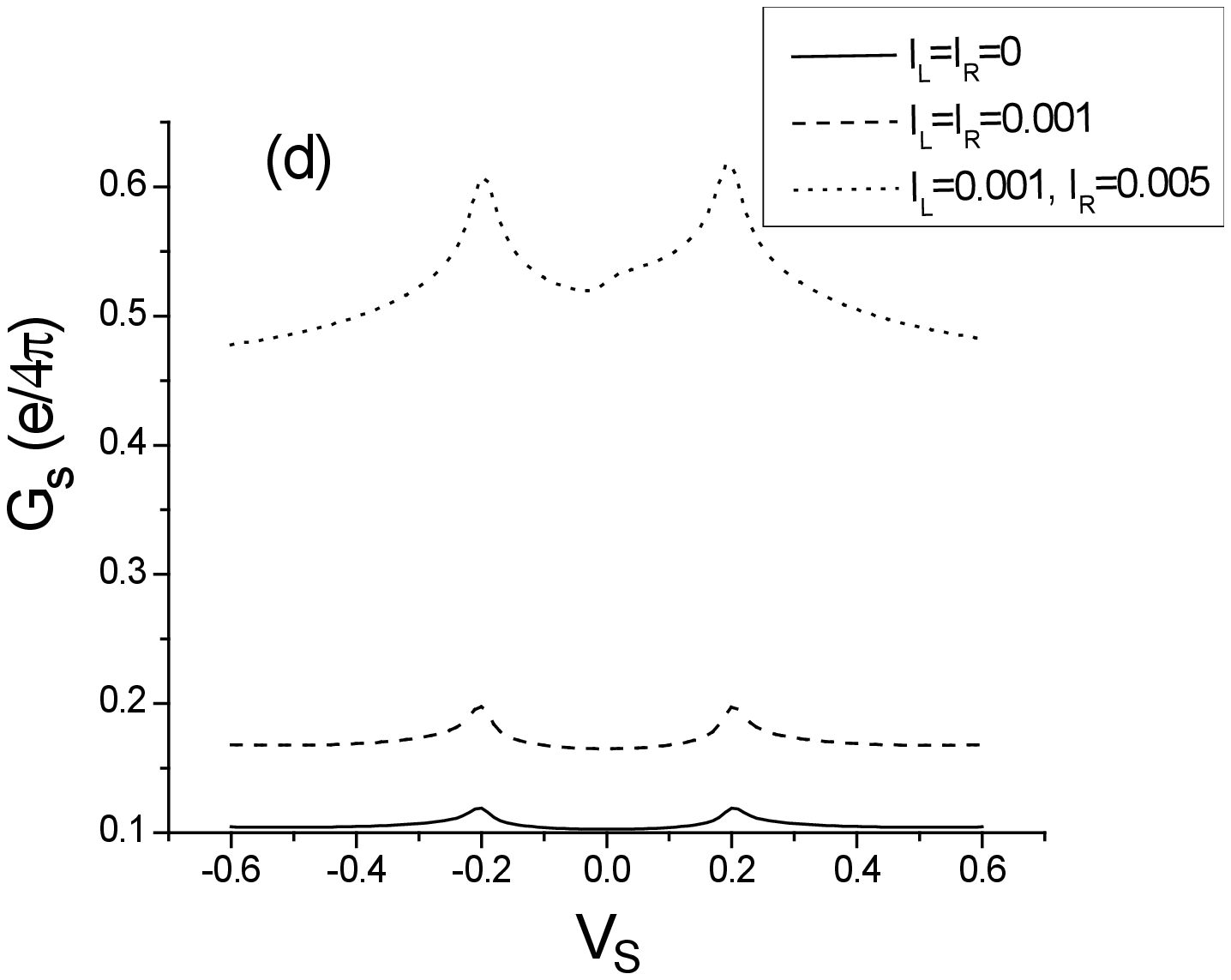}
\caption{\label{fig:currentjie2} (a)-(b) Schematic energy diagrams
of the Kondo density-of-state peaks and the electrochemical
potentials of the two leads, for the case of
$\Delta\varepsilon_\uparrow>V_s$ and
$\Delta\varepsilon_\uparrow<V_s$. (c)-(d) The differential charge
and spin conductance $G_c$ and $G_s$ plotted as a function of the
spin bias $V_s$ for different lead-dot interaction strengths. Solid,
dashed and dotted curves correspond to the case of $I_L=I_R=0$,
$I_L=I_R=0.001$ and $I_L=0.001, I_R=0.005$, respectively. The
energies of the quantum dot are chosen to be
$\varepsilon_\uparrow=-2.9$, $\varepsilon_\downarrow=-3.1$, so that
$\Delta\varepsilon_\uparrow=0.2$. Other parameters are the same as
those of Figure~\ref{fig:midu1}.}
\end{center}
\end{figure}
As shown in Figure~\ref{fig:midu1}, the spin-up and spin-down
density of states are plotted in the presence of dot-lead Coulombic
interactions described by $I_L$ and $I_R$. In the absence of the
dot-lead interaction, i.e., $I_L=I_R=0$ [solid line], a broad main
peak is observed at $\epsilon\sim -2.5$, which is associated with
the renormalized level $\varepsilon_\uparrow$ of the quantum dot. In
addition, there are two sharp Kondo peaks at energies
$\epsilon\approx 0$ and $\epsilon\approx 0.35$ [see
Figure~\ref{fig:midu1}(a)]. The Kondo peaks for spin $\sigma$ arise
from the contribution of the self-energy $\Sigma'_{2\sigma}$, due to
virtual intermediate states in which the site is occupied by an
electron of opposite spin $\bar{\sigma}$ \cite{Meir2}. The real part
of $\Sigma'_{2\sigma}$ grows logarithmically near the energies
$\epsilon_{\sigma}=\mu_{\alpha\bar{\sigma}}+\Delta
\varepsilon_\sigma$ with $\Delta
\varepsilon_\sigma=\varepsilon_\sigma-\varepsilon_{\bar{\sigma}}$,
due to the sharp Fermi surface at low temperature. This logarithmic
increase translates into peaks in the density of states near those
energies. According to the same physical explanation, we can deduce
that the Kondo peaks of the spin-down density of state
$\rho_\downarrow$ should occur at energies $\epsilon=-0.05$ and
$\epsilon=-0.35$, as can be confirmed from
Figure~\ref{fig:midu1}(b). In the presence of the dot-lead
interaction, for example $I_L=I_R=0.001$ (dashed lines), the
position of the broad main peak is shifted. However, the positions
of the Kondo peaks are not affected by the dot-lead interaction
which induces the self-energies $\Sigma_{3\sigma}$ and
$\Sigma_{4\sigma}$ and do not contribute to the Kondo effect. When
the interaction strength is increased, e.g., $I_R=0.005$ (dotted
line), the position of the broad main peak is shifted further to
higher energy compared to the previous case, but the positions of
the two Kondo peaks remain invariant.

Next, we investigate the effect of the dot-lead interaction on the
charge and spin differential conductances, which are defined as
$G_c=dJ/dV_s$ and $G_s=dJ_s/dV_s$, where
$J_s=\hbar(J_{\uparrow}-J_{\downarrow})/2e$, respectively. The
differential conductance $G_c$ and $G_s$ are plotted as a function
of the spin bias $V_s$ for different interaction strengths as shown
in Figure~\ref{fig:currentjie2}. To explain the observed trends in
the spin and charge differential conductance, we sketch the
electrochemical potentials in the two leads, and superimpose on it
the Kondo peaks in the density-of-states [see
Figs.~\ref{fig:currentjie2}(a) and (b)]. We observe a plateau in
$G_c$ over the bias interval of $|V_s|\leq 0.2$, for the cases of
$I_L=I_R=0$ and $I_L=I_R=0.001$. However, the conductance plateau is
destroyed in the case of asymmetric lead-dot interaction, i.e.
$I_L\neq I_R$, and the charge differential conductance assumes a
negative value over the same bias interval [see
Figure~\ref{fig:currentjie2}(c)]. The spin differential conductance
$G_s$ shows two Kondo peaks at $|V_s|= 0.2$, irrespective of the
symmetry or strength of the lead-dot interactions [shown in
Figure~\ref{fig:currentjie2}(d)]. $G_s$ is also enhanced with
increasing strength of the dot-lead interactions.
\begin{figure}
\begin{center}
\includegraphics[width=5.5cm]{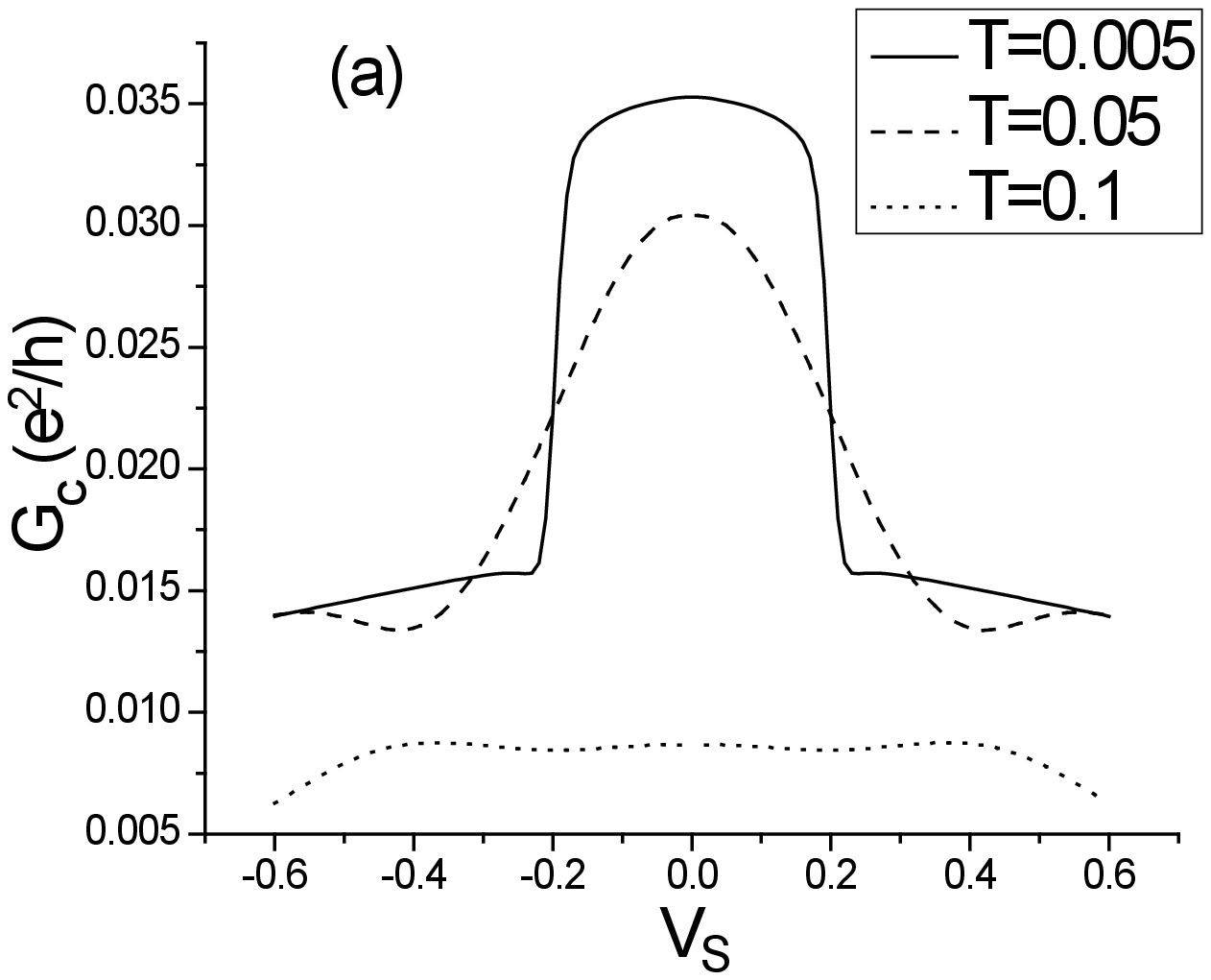}
\includegraphics[width=5.5cm]{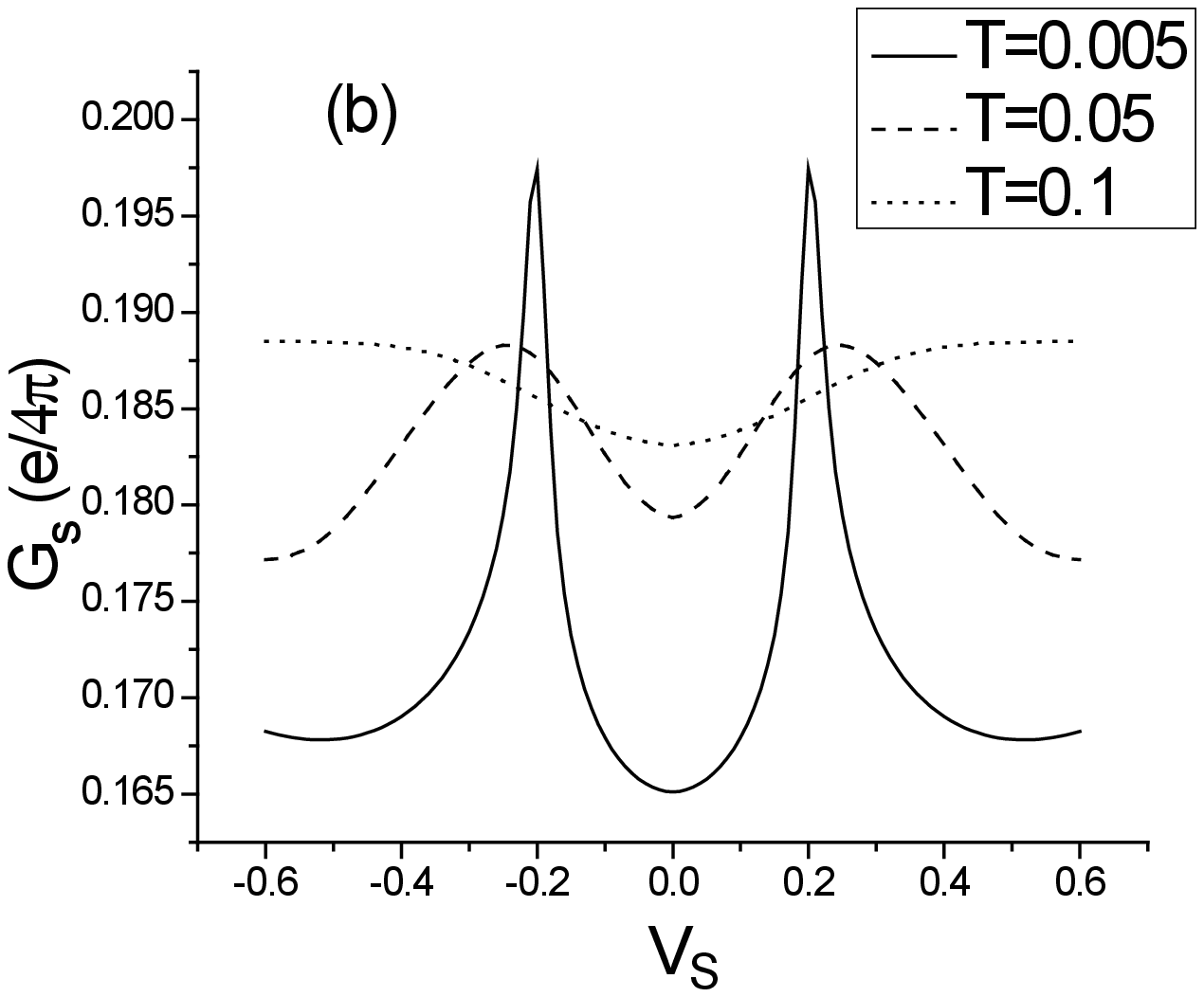}
\caption{\label{fig:conductance2}  The differential charge and spin
conductance $G_c$ (a) and $G_s$ (b) versus the spin bias $V_s$, for
temperatures $T=0.005$ (solid line), $T=0.05$ (dashed line) and
$T=0.1$ (dotted line). The lead-dot interaction strength is set at
$I_L=I_R=0.001$. Other parameters are the same as those of
Figure~\ref{fig:midu1}.}
\end{center}
\end{figure}
We note that the spin-up and spin-down electrons flow along opposite
directions, i.e., $J=|J_{\uparrow}|-|J_{\downarrow}|$ and
$J_s\propto|J_{\uparrow}|+|J_{\downarrow}|$, and that the two
currents $J_{\uparrow}$ and $J_{\downarrow}$ can be different due to
the energy splitting $\Delta\varepsilon_\uparrow$. To explain the
above conductance dependence on $V_s$, we refer to the  schematic
diagram of the Kondo density-of-state peaks [see
Figure~\ref{fig:currentjie2}(a)-(b)]. The Kondo peaks begin to enter
the spin-bias conduction window when the spin bias is increased
beyond $|V_s|=\Delta\varepsilon_\uparrow$. This results in the spin
differential conductance $G_s$ having two Kondo peaks at
$V_s=\pm\Delta\varepsilon_\uparrow$. The entry of the two Kondo
peaks into the conduction window also reduces the difference in the
magnitude of $J_\uparrow$ and $J_\downarrow$, thus resulting in a
sharp drop (plateau step) in the charge conductance at
$V_s=\pm\Delta\varepsilon_\uparrow$. The conductance plateau can
thus be attributed to the combined effect of the spin bias in the
leads and the Zeeman splitting in the QD. The dot-lead interaction
tends to increase the coupling between the leads and the QD, thus
resulting in a general increase in the charge and spin differential
conductances. In the presence of asymmetrical dot-lead interaction,
i.e., $I_L=0.001$ and $I_R=0.005$, the symmetry in the transport
across the QD is broken, and thus the conductance plateau
disappears. Two conductance dips occur at
$|V_s|<\Delta\varepsilon_\uparrow$ due to the contribution from the
Kondo peaks in the density-of-states.

Finally we study the temperature dependence of the differential
conductances. As shown in Figure~\ref{fig:conductance2}, the two
Kondo peaks in $G_s$ at $V_s=\pm 0.2$, and the conductance plateau
in $G_c$ in the bias interval $|V_s|<0.2$ can be clearly observed
at a low temperature of $T=0.005$. With increasing temperature,
e.g., at $T=0.05$, the Kondo peaks become thermally broadened, while
the plateau in $G_c$ sharpens into a peak profile. With a further
increase in temperature to $T=0.1$, the plateau in $G_c$ is almost
completely suppressed. These changes can be largely attributed to
the thermal distribution of electrons about the electrochemical
potential in the leads. The thermal distribution in turn affects the
self-energy $\Sigma_{2\sigma}'$ of the intra-dot Coulomb
interaction, which is primarily responsible for the Kondo resonances
in the density-of-states [see Eq.~(\ref{equation9a})].

\section{Summary}\label{summary}
In this work, we analyze the spin-transport properties of a quantum
dot system driven by spin bias in the presence of dot-lead Coulombic
interactions. The transport property is discussed on the basis of
Keldysh nonequilibrium Green's function framework. According to the
equation-of-motion technique and Langreth's theorem, we derive the
analytical expression of the current through the quantum dot in the
presence of the dot-lead Coulombic interaction. Our numerical
results show that although the interaction can renormalize the
energy levels of the quantum dot, they leave the position of the
Kondo peaks in the density of states unchanged. This is because the
Kondo effect arises primarily for intra-dot Coulomb interactions
involving electrons of opposite spins. The Kondo resonances in the
density of states translate into peaks in the spin differential
conductance when the magnitude of the spin bias is equal to that of
the Zeeman energy split in the quantum dot. There also exists a
plateau in the charge differential conductance at low bias, due to
the combined effect of spin bias and the Zeeman energy splitting.
The position of the steps of the conductance plateau can also be
attributed to the Kondo effect. The strength of the Coulombic
lead-dot interactions affects the magnitude of both the spin and
charge conductances. Furthermore, in the presence of asymmetrical
dot-lead interaction strengths, the plateau in the charge
conductance disappears, and is replaced by conductance peaks.
Finally, the temperature dependence of the differential conductances
is qualitatively discussed.

The authors would like to thank the Agency for Science, Technology
and Research (A*STAR) of Singapore, the National University of
Singapore (NUS) Grant No. R-398-000-061-305 and the NUS Nanoscience
and Nanotechnology Initiative for financially supporting their work.
The work was also supported by Innovation Research Team for
Spintronic Materials and Devices of Zhejiang Province.
\bibliographystyle{model1-num-names}

\end{document}